\documentclass[]{raa}            % referee version: for submission

\usepackage{graphicx,times}             %for PS/EPS graphics inclusion, new
\usepackage{natbib}
\usepackage{amssymb,amsmath}
\usepackage{mathtools}
\usepackage{subcaption}

\bibpunct{(}{)}{;}{a}{}{,}

\usepackage[colorlinks=true, allcolors=blue]{hyperref}

\begin{document}

   \title{Cross-method analysis of co-rotation radii dataset for spiral galaxies}

 \volnopage{ {\bf 20XX} Vol.\ {\bf X} No. {\bf XX}, 000--000}
   \setcounter{page}{1}

   \author{V. S. Kostiuk \inst{1}
   \and A. A. Marchuk \inst{2,3} 
   \and A. S. Gusev \inst{1}
   }

   \institute{
Sternberg Astronomical Institute, Lomonosov Moscow State University, Universitetsky pr. 13, 119234 Moscow, Russia; {\it valeriekostiuk@yandex.ru}\\
        \and
             Central (Pulkovo) Astronomical Observatory, Russian Academy of Sciences, Pulkovskoye chaussee 65/1, 196140 St. Petersburg, Russia\\
	\and
	 Saint Petersburg State University, Universitetsky pr. 28, 198504 St. Petersburg, Russia\\
\vs \no
   {\small Received 20XX Month Day; accepted 20XX Month Day}
}

\abstract{A co-rotation radius is a key characteristic of disc galaxies that is essential to determine the angular speed of the spiral structure $\Omega_p$, and therefore understand its nature. In the literature, there are plenty of methods to estimate this value, but do these measurements have any consistency? In this work, we collected a dataset of corotation radius measurements for 547 galaxies, 300 of which had at least two values. An initial analysis reveals that most objects have rather inconsistent corotation radius positions. Moreover, a significant fraction of galactic discs is distinguished by a large error coverage and almost uniform distribution of measurements. These findings do not have any relation to spiral type, Hubble classification, or presence of a bar. Among other reasons, obtained results could be explained by the transient nature of spirals in a considerable part of galaxies. We have made our collected data sample publicly available, and have demonstrated on one example how it could be useful for future research by investigating a winding time value for a sample of galaxies with possible multiple spiral arm patterns.
}

   \authorrunning{V. S. Kostiuk et al. }            %author_head in even pages
   \titlerunning{Cross-method analysis of co-rotation radii dataset}  % title_head in odd pages
   \maketitle

%________________________________________________ sections below
% 
\section{Introduction}           %% first-level sections will be auto-capitalized
\label{sect:intro}
The nature and evolution of the spiral structure of disc galaxies is still a question that is not well-understood, despite the long time of investigation. There are two opposing points of view: long-lived stationary and dynamic (transient or recurrent) spirals. The first one, based on the so-called quasi-stationary density wave theory~\citep{Lin&Shu}, maintains that spiral arms are waves of increased density that propagate through regions where stars and gas gather in the galaxy's disc. The visual representation of this theory was given in~\citet{Kalnajs73}. According to this paper, when stars' orbits are rearranged in a certain coordinate system, their apocentres take the shape of spiral arms (see fig.~3 in~\citealp{Kalnajs73}). Thus, due to the slower movement of the stars near the apocentres, the material in these areas slows down and accumulates. The angular speed of the spiral structure, $\Omega_p$ (a frequency of stellar orbit's precession), is supposed to be constant with galactocentric radius. The position where the angular speeds of the disc and the pattern become equal, called corotation radius ($R_c$) or corotation resonance. In contrast, the transient spirals' theory assumes that the spiral structure has a dynamic nature, and its angular speed value changes the same way as the angular velocity of disc rotation \citep{Baba13,Sellwood10}. Note that in this case, spirals can seem to be long-lived, although they are actually composed of many segments that are most often torn apart and reconnected to other segments of spiral arms \citep{Fujii11}.
\par
Despite the persuasiveness of the theories mentioned above, each of them has a number of advantages and disadvantages. The long-lived spiral structure existence is provided that the effects amplifying the wave moving in the radial direction at resonances are taken into account (swing amplification, see~\citealp{Toomre81}). Besides, the classical density wave theory based on tight-winding approximation (short wavelength or WKB\footnote{Named after the Wentzel-Kramer-Brillouin approximation in quantum mechanics.} approximation) does not explicate the existence of spiral structure with large pitch angles\footnote{Pitch angle --- the angle between the tangent to the spiral arm and the line perpendicular to the radius vector directed from the point to the galactic centre.} and big number of arms \citep{LinShu1967}. In order to solve these problems, the initial representation of the density wave spiral had been transformed into the theory of global modes (see~\citealp{Bertin83,BertinLin(1996)}, for example). As for the dynamic spiral pattern, it is much easier to form in numerical simulations. This structure is considered to be generated by gravitational instability caused by the swing amplification mechanism \citep{SellwoodCarlberg(1984)}. Compared to stationary spiral density waves, which formation and maintenance occur only under special conditions \citep{D'Onghia2013}, recurrent spirals are generated at any configurations of the disc parameters, even in the presence of weak instability. However, the mechanisms promoting continuous regeneration of spiral structures are still not established (see reviews~\citealp{Dobbs&Baba,Sellwood&Masters} for more details).
\par
Also, it is worth to mention another theory based on the interaction with satellites. The tidal forces from a companion not only form various structures in a disc (bridges and tails,~\citealp{Toomre72}) but also influence the formation of spiral structure~\citep{Donner94}. However, it is not yet entirely clear what spirals are generated, whether they are stationary or dynamic. In addition, some studies investigate the influence of bars, which are often connected with the tips of spiral arms. For example, \citet{Athanassoula10} found that a bar has an impact on the pitch angle of the spirals. However,~\citet{Rautiainen99} showed that the bar and the spirals can be independent features% do not affect each other
and rotate with different angular velocities. Besides, some researchers maintain that spirals can be formed due to the nonlinear coupling between bars and spiral density waves~\citep{Minchev12}.
\par 
We cannot omit that galaxies can have multiple spiral modes rotating at different angular velocities~\citep{Efthymiopoulos20,Meidt08}. These modes are connected to each other through resonances. The multiplicity of $\Omega_p$ values means that there are several positions of the corotation radius, which has been confirmed in observations \citep{Buta&Zhang(2009),F-B14}. Additionally, there are several models in which each spiral mode has a unique angular velocity (see fig.~8 in~\citealp{Forgan18}).
\par
In practice, it is quite difficult to determine which theory better describes the observed spirals in a galaxy, and it may require a combination of such models. From an observational perspective, it is not easy to distinguish between long-lived and transient spiral arms, but at the moment, there are more and more observational tests that can testify to each of the theories. Most of these tests are based on the analysis of the spiral pattern angular velocity, the constant profile of which indicates the existence of a quasi-stationary density wave, and if it changes with distance, then the spirals are more likely to have a recurrent origin. In addition to this, there are other indirect signs related to the angular frequency, such as the age gradient of stars across the arm~\citep{Puerari97,Tamburro08,Egusa09}, morphological features near resonances \citep{Elmegreen(1992)}, and others. Most often, to find the pattern angular speed $\Omega_p$, the velocity curve and the corotation radius location are used. There are plenty of methods applied for $R_c$ value determination (the number of which continues to grow to this day, for example,~\citealp{Pfenniger23, Marchuk2024}).

\par
To sum up, one of the essential problem in galactic dynamics relates to the nature of spiral structure, and the estimation of corotation radii plays a crucial role in solving this issue. Various methods have been developed to measure its position, but all of them rely on different arguments and assumptions. However, the use of different methods can lead to contradictory results. For example,~\citet{ScaranoLepine(2012)} showed that part of their sample of objects had consistent $R_c$ defined by several methods, while the others demonstrated disagreement in estimates of corotation resonances. What could cause such a discrepancy? Firstly, some objects may have multiple spiral modes, leading to several $R_c$ values. Additionally, the obtained results may testify in favour of dynamic spirals' theory, according to which there is no localised corotation resonance position within the galactic disc. Furthermore, the assumption of some unreliable methods and their measurements cannot be excluded. Hence, the primary goal of this paper is to explore the possible explanations for the discrepancy in $R_c$ measurements.
\par
It is necessary to emphasize that the estimation of the corotation radius position is also crucial for solving other astrophysical problems. For instance, using an inaccurate corotation radius value and the corresponding pattern angular speed can lead to unreliable results. \citet{Spurring} investigated the star formation processes in the spurs of NGC~628, and found that the stars are forming in-situ within the spurs, and do not drift out of the spiral arms. To reach this conclusion, the researchers used the angular speed value obtained using Tremaine-Weinberg method (see fig.~15 in \citealp{Williams(2021)}). However, in this figure a linear regression embedded within a dispersed point cloud, and there are other $\Omega_p$ measurements available in the literature and our sample, which are not fully consistent with each other. This raises question about the accuracy of the angular velocity determination and consequent implications in~\citet{Spurring}. Furthermore,~\citet{Spitoni2023} demonstrated the influence of the corotation resonance on the chemical evolution in a recent paper. Moreover, a swing amplification mechanism is believed to occur near the $R_c$ position, allowing the spirals to change their direction of rotation and amplify there \citep{Goldreich1978,Toomre81}. The corotation resonance position also has a direct connection to the transfer of angular momentum in the galactic disc, which is essential to understand its evolution~\citep{SellwoodBinney02}. In summary, the corotation radius is a resonance $1 : 1$, playing a key role in many dynamic processes, and its examination can shed light on the nature of the spiral structure.
\par
To investigate the posed question, we collected a dataset of corotation radius measurements, and its description is given in Section~\ref{sect:sample}. In Section~\ref{sect:analysis}, we performed a statistical analysis on the collected sample and constructed distributions of the corotation radii for certain objects. In Section~\ref{sect:discussion}, we obtained a total coverage of the corotation radii errors and a measure of \textit{consistency} for each galaxy. We examined possible reasons for the appearance of its large values, as well. Section~\ref{sect:wind_time} is dedicated to the consideration of galaxies that may have multiple spiral modes. Moreover, we compared winding and rotating times for this set of objects. In Section~\ref{sect:conclusion} we summarize the main results and discuss the potential application of the collected data for future research. A brief summary of the methods used for measuring the corotation resonance positions is presented in the Appendix~\ref{sect:methods}.

\section{Corotation radii data}
\label{sect:sample}
The main goal was to collect a sample of corotation radius values, obtained using conceptually different methods, and to find out whether they were consistent with each object. If not, we needed to analyse potential reasons for this discrepancy. As a result, we collected corotation radius positions from 47 research papers in which these values had been directly measured. It is important to note that we did not generally pay attention to papers containing $R_c$ values for a single object or several objects (\citealp{Sempere1997,Seigar2018}); instead, we focused on papers with a relatively large number of galaxies and measurements (for example,~\citealp{Buta&Zhang(2009),Cuomo20,ElmegreenElmegreen(1995),Sierra2015,Williams(2021)}).
\par
Hence, our dataset consists of~1711 corotation resonances for~547 galaxies, which were estimated using~12 methods. An example of a collected sample is presented in Table~\ref{tab1}. This table contains the name of the object, corotation radius with errors, the method used, band, optical radius, the reference for the paper from which the measurement was taken, and a link to the original source. If necessary, we have converted the $R_c$ value to arcseconds by using the galaxy's distance specified in the relevant article. Note that methods applied for corotation radii estimation were divided into conditional groups, names of which are contained in the table.
\textit{P-D} and \textit{offset} methods are based on the assumption that the angular offset between young and old stellar population changes a sign at the vorticity of corotation \citep{Puerari97,Tamburro08}. \textit{Width} method analyses the azimuthal distribution of matter along the spiral arms \citep{Marchuk2024}. \textit{T-W} enables us to obtain the pattern speed directly \citep{T-W}, while in the \textit{model} method this value varies until the simulated galaxy has the same features as the observable one. \textit{Potential-density} and \textit{metallicity} methods use the radial profiles of the phase shift \citep{Buta&Zhang(2009)} and metallicity \citep{ScaranoLepine(2012)}. $R_c$ locations of gathered sample were also estimated by analysing the residual velocities (\textit{F-B}: \citealp{Font11,F-B14}) and morphological features (\textit{morph}: \citealp{Elmegreen(1992),ElmegreenElmegreen(1995)}) of galaxies. More detailed descriptions of listed methods are presented in the Appendix~\ref{sect:methods}. Column (5) shows what kind of data was applied to measure the corresponding $R_c$ value. 
Also, the advanced version of our dataset includes information about the component (stellar or gaseous) used to determine each measurement, however it is worth to take into account the subjective content of this column. These data were collected to investigate correlations and systematic errors in the case of inconsistent measurements.
Additionally, this dataset includes corotation radii for bar structures. One of the potential directions of this research may lie in investigating the relationship between the bar and spirals rotation speeds. It was not our primary goal, consequently this dataset contains only a small portion of the available data about bars' $R_c$, however, it will be updated in the future. In this work, we aim to explore only the corotation resonances related to the spiral structure, and therefore we did not concentrate on methods related to bar features (in our notation `bar-torque':~\citealt{Verley07}; `rings':~\citealt{Perez12}; `gaps':~\citealt{Buta17}). As mentioned previously, a number of these measurements is relatively small, and their presence did not significantly influence the following results in this paper. 

A complete dataset is publicly available at GitHub.\footnote{\url{https://github.com/ValerieKostiuk/CRs\_dataset}} As mentioned before, the collected sample does not include absolutely all the $R_c$ measurements available in the literature. We plan to update it over time and we would greatly appreciate any help. This data will be valuable for researchers developing and applying different methods of corotation radii estimation in order to compare determined values with those taken from literature. Moreover, the gathered dataset enables us to determine a reliable $R_c$ value for a certain part of objects. This quantity is not only substantial for estimating the pattern angular speed, but its position plays a significant role in chemical and dynamical evolution in galaxies. In this paper, we focused only on analysis of the gathered dataset. Specifically, we reviewed the measurement data for consistency and considered several factors that could affect on the results obtained in the following sections. The code used for analysis is also available at GitHub. 

\par
\renewcommand{\arraystretch}{1.4}
\begin{table}[ht]
\bc
\begin{minipage}[]{130mm}
\caption[]{This table presents the corotation radius measurements (column 3) for a sample of galaxies (column 1) which were determined by different methods (column 4). Columns~6 and~5 shows observational data type and a reference of paper from which $R_c$ value were taken. Also, there is the optical radius (column 2) of the galaxy according to NED.\label{tab1}}\end{minipage}
\setlength{\tabcolsep}{3pt}
\small
 \begin{tabular}{l|r|r|l|l|l}
  \hline\noalign{\smallskip}
Galaxy name& $r_{25}$, arcsec & $R_c$, arcsec &Method& Band& Reference\\
\hline\noalign{\smallskip}
(1)& (2) & (3) & (4)& (5)& (6)\\
  \hline\noalign{\smallskip}
IC 0342& $641.4$ & $344.00 ^{+26.0}_{-79.0}$ & T-W & $\text{H}_{\alpha}$ & ~\citet{Fathi}\\
NGC 0613 & $164.9$ &  $126.20 ^{+14.6}_{-14.6}$ & model & H &            ~\citet{RautiainenLaurikainen(2008)} \\
NGC 0895 & $108.9$ & $60.00 ^{+4.8}_{-4.8}$ & morph &  optical &          ~\citet{ElmegreenElmegreen(1995)}           \\
NGC 1097 &$280.0$ & $96.60 ^{+30.5}_{-30.5}$&T-W &CO,$\text{H}_{\alpha}$& ~\citet{Williams(2021)}        \\
NGC 1365 &$366.6$ &$229.93 ^{+59.7}_{-59.7}$&metallicity&$12$+$\log(\text{O}/\text{H})$	&    ~\citet{ScaranoLepine(2012)}                  \\
NGC 1566 &$249.6$ &$122.17 ^{+45.4}_{-45.4}$&offset&B, Spitzer, GALEX	&    ~\citet{Abdeen20}       \\
NGC 2403&$656.4$ &$392.32 ^{+35.6}_{-35.6}$&metallicity&$12$+$\log(\text{O}/\text{H})$	&    ~\citet{ScaranoLepine(2012)}                  \\
NGC 2903&$377.7$ &$122.43 ^{+32.3}_{-32.3}$&offset&B, Spitzer, GALEX	&    ~\citet{Abdeen20}                  \\
NGC 3631 & $150.4$ & $63.60 ^{+5.4}_{-5.4}$ & morph &  optical &          ~\citet{ElmegreenElmegreen(1995)}           \\
NGC 3686 & $52.2$ & $35.6 ^{+5.1}_{-5.1}$ & model &  H &          ~\citet{RautiainenLaurikainen(2008)}          \\
NGC 4536 &$227.5$& $113.65 ^{+27.5}_{-27.5}$&T-W &CO,$\text{H}_{\alpha}$& ~\citet{Williams(2021)}        \\
NGC 5033 &$321.5$ &$158.35 ^{+5.4}_{-5.4}$&offset&B, Spitzer, GALEX	&    ~\citet{Abdeen20}                  \\
NGC 5248& $185.0$ & $102.71 ^{+0.0}_{-0.0}$ & morph &  B &          ~\citet{Elmegreen(1992)}           \\
NGC 5364 & $202.8$ & $113.4 ^{+3.6}_{-3.6}$ & morph &  optical &          ~\citet{ElmegreenElmegreen(1995)}           \\
NGC 5427 &$85.6$& $64.5 ^{+0.9}_{-0.9}$&F-B&HI,$\text{H}_{\alpha}$&~\citet{F-B14}   \\
NGC 7552 & $101.7$ &  $65.00 ^{+5.9}_{-5.9}$ & model & H &            ~\citet{RautiainenLaurikainen(2008)} \\
MESSIER~066 &$273.6$&$163.00^{+0.0}_{-0.0}$&T-W&CO	&    ~\citet{RandWallin(2004)}                  \\
MESSIER~074&$314.2$&$88.8^{+4.2}_{-4.2}$&morph&optical	&    ~\citet{ElmegreenElmegreen(1995)}                  \\
MESSIER~091&$161.1$&$21.91^{+0.9}_{-0.9}$&P-D&SDSS	&    ~\citet{Sierra2015}                  \\
MESSIER~099&$161.0$&$77.35^{+11}_{-11}$&model&NIR, H$_\alpha$	&    ~\citet{Kranz03}                  \\
MESSIER~100&$222.4$& $147.90 ^{+10.7}_{-10.7}$&F-B&HI,$\text{H}_{\alpha}$&~\citet{F-B14}     \\            
  \noalign{\smallskip}\hline
\end{tabular}
\ec
\end{table}

\section{Corotation radii data analysis}
\label{sect:analysis}
In order to understand if there is any consistency in 
determination of the corotation resonance position, it is necessary to have more than one $R_c$ value for each object. It is important to note that 386 of the galaxies in this sample only have measurements of $R_c$ that were determined using one method. For 247 of these galaxies, only one $R_c$ value was found. A reliable determination of the corotation position requires at least two methods to be applied to the same object. This requirement is met by the remaining 161 galaxies in the sample.
\par
\begin{figure}
   \centering
  \includegraphics[width=10.5cm, angle=0]{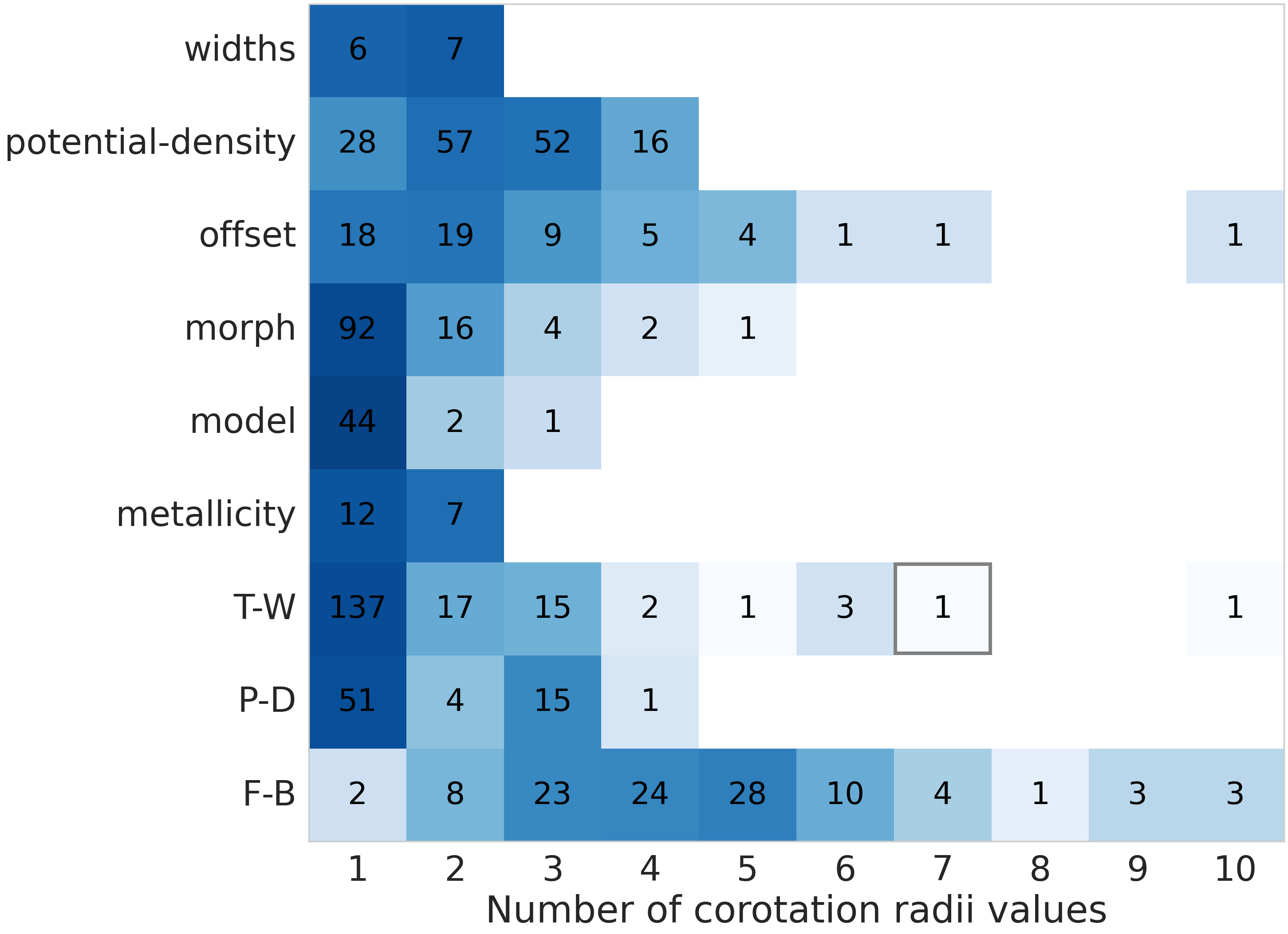}

   \caption{A two-dimensional histogram of the distribution of the number of corotation radius measurements (x-axis) depending on the method (y-axis) used to estimate them. The number in each cell determines the number of galaxies for which a certain number of $R_c$ values were found using a particular method. The colour gradient shows the change in the number of objects. See the text for details.} 

\label{method_cr_num}
   \end{figure}

Figure~\ref{method_cr_num} shows how many measurements determined by a particular  method are presented in the collected sample. For example, a cell marked by grey square means that there is only one galaxy (M~100) in the sample for which 7 different corotation radius values were estimated by \textit{T-W} method taken from several research papers (\citealp{Hernandez05,Williams(2021),RandWallin(2004)}). According to this figure, the \textit{F-B} and \textit{potential-density} methods measure most frequently for more than one corotation resonance value.
The presence of several $R_s$ measurements found by these methods may be associated with the existence of multiple spiral modes in the galaxy~\citep{Buta&Zhang(2009)}. Note, that the \textit{F-B} method can also measure the positions of resonances that may not be related to corotation ones (see table~7 in \citealp{F-B14}), which are also included in our dataset. Therefore, in order to distinguish the most plausible corotation from other measurements, it is necessary to compare the locations of this method to that obtained by other ways.

\par
\begin{figure}
   \centering
  \includegraphics[width=12.5cm, angle=0]{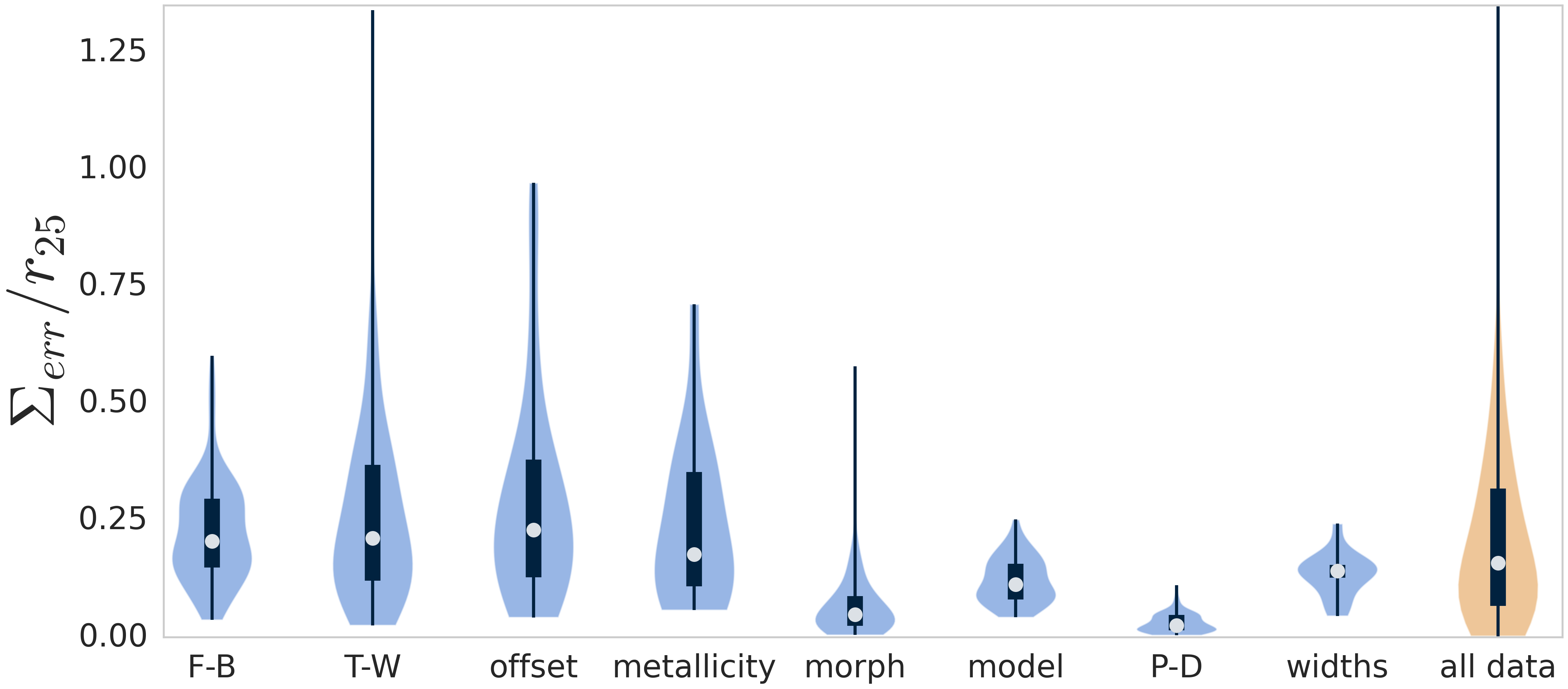}

   \caption{Violin plots of the distribution of the total error coverage in fractions of the optical radius $\Sigma_{err}/r_{25}$ for methods that provide a non-zero error in determining the corotation radius. The rightmost diagram shows the distribution of the corresponding quantity for the entire dataset.} 

\label{method_err_violin}
   \end{figure}

Before understanding whether there is any agreement between the measurements of $R_c$ determined by different methods, it is necessary to examine their positions in each galaxy, measured with the same method. In order to do this, we calculate the total coverage of the corotation radii error ($\Sigma_{err}$) for each method. If the values of $R_c$ intersect within the limits of error, then $\Sigma_{err}$ consists of joining their errors. If they do not intersect, then the error covers add up. The magnitude of $\Sigma_{err}$ directly relates to the range of angular pattern speed $\Omega_p$ value estimation. To compare the total error coverage magnitude between galaxies of different sizes, we normalize this value by the optical radius $r_{25}$. As a result, violin diagrams were constructed. Figure~\ref{method_err_violin} shows the distribution of the value $\Sigma_{err}/r_{25}$ for each method (except \textit{potential-density}, which in the original paper did not estimate $R_c$ uncertainties). In Fig.~\ref{method_err_violin}, it can be seen that the median value of the total error coverage does not exceed a quarter of the optical radius of the galaxy. However, there are a number of galaxies for which the error of corotation radius value determined using the Tremaine-Weinberg and the age gradient methods is more than $75$\% of the entire disc, and sometimes exceeds the value of the optical radius. As for the Font-Beckman method, a large value of $\Sigma_{err}$ is associated with its ability to estimate not only the corotation resonances, which can be distributed evenly across the disc. The error in measuring $R_c$ by the metallicity gradient method can be associated with the sparsity of the data, making it impossible to accurately estimate its position. Figure~\ref{method_cr_num} demonstrates a relatively small number of galaxies in which the Tremaine-Weinberg method has measured more than one corotation radius. Consequently, the magnitude of $\Sigma_{err}$ for this method is explained rather by a large error bar of individual values of the corotation radius, rather than by their inconsistency.

\par
\begin{figure}
   \centering
  \includegraphics[width=12.5cm, angle=0]{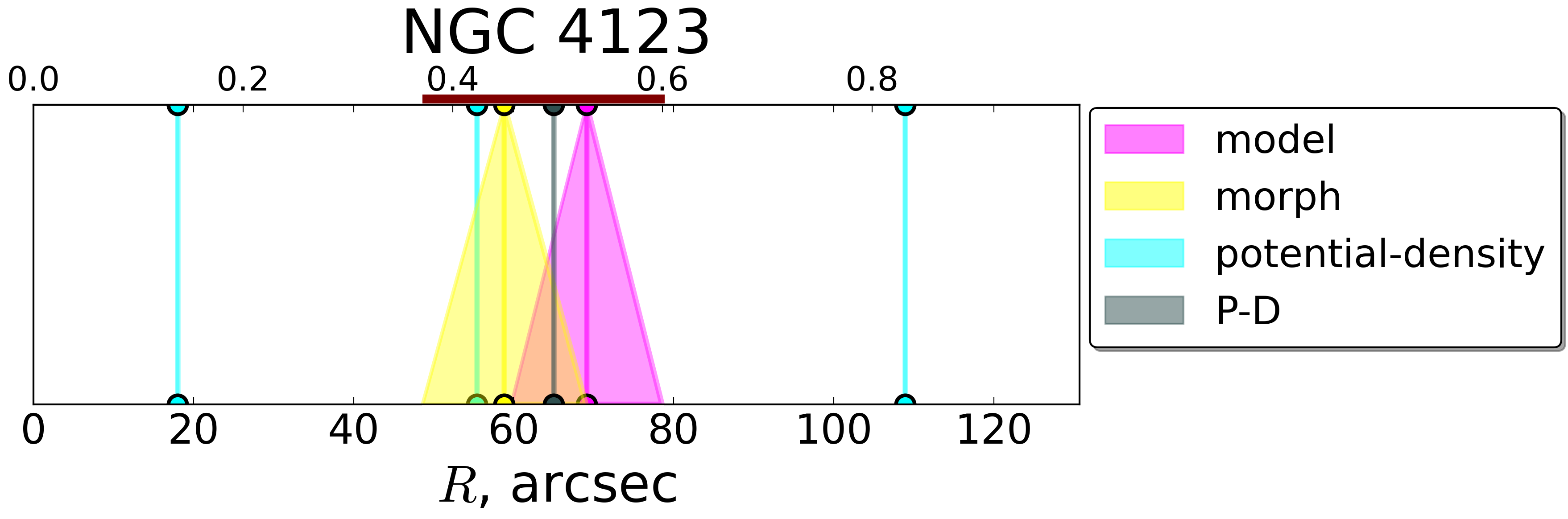}

   \includegraphics[width=12.5cm, angle=0]{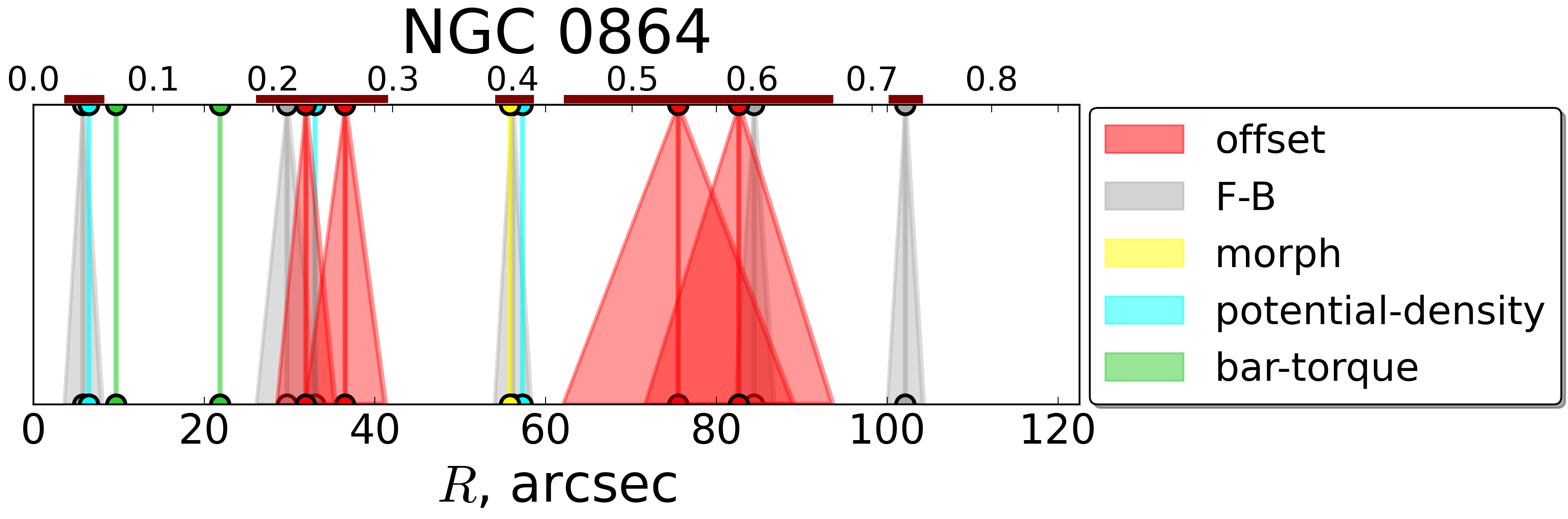}

    \includegraphics[width=12.5cm, angle=0]{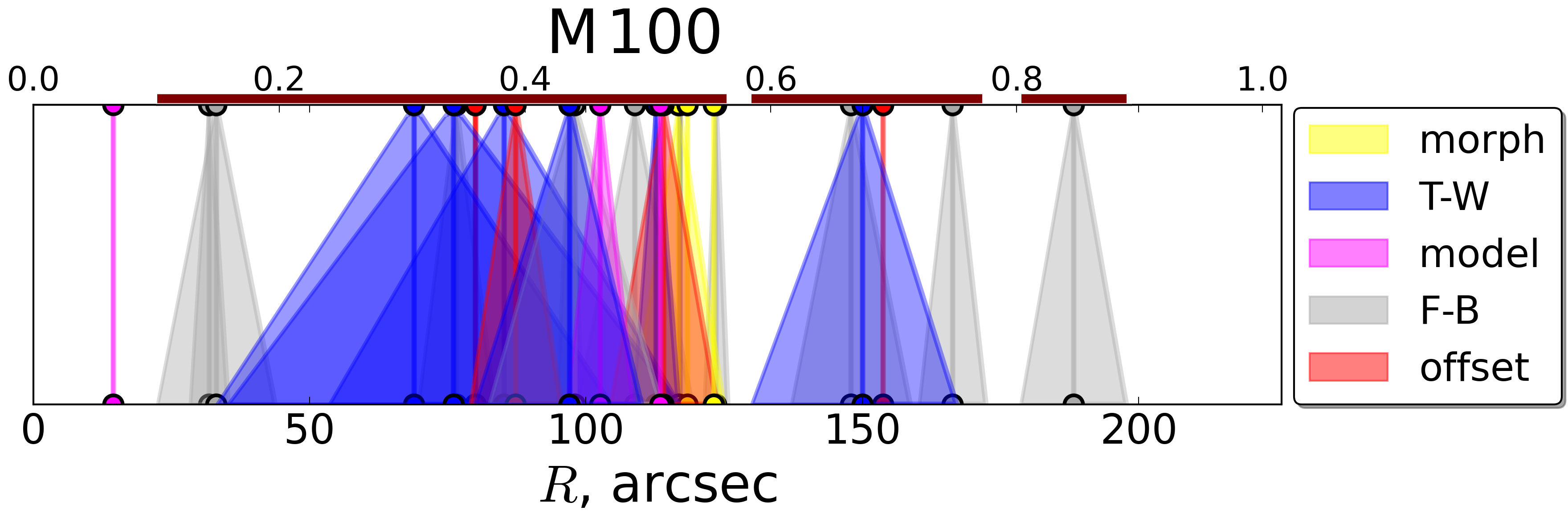}

   \caption{The distribution of corotation radii (vertical lines) for the galaxies NGC~4123 (top), NGC~864 (middle), and M~100 (bottom). The shaded areas show the error in determining each value (base of the cone). Their colour indicates the method used (see legend on the right). The lower scale indicates their positions in arcseconds, and the upper scale indicates their positions in units of the optical radius. Brown horizontal lines on top of each plot mark the areas of errors joining, which make up the quantity of $\Sigma_{err}$.} 

\label{distribution}
   \end{figure}

To investigate whether the positions of corotation resonances in the same galactic disc are consistent, we examined the distributions of  their values for those objects\footnote{Full sample of images with such distributions are presented in online catalogue.} whose  measurements were estimated by different methods (see  Fig.~\ref{distribution}). We visually viewed each distribution and found that only 15\% of the 161 galaxies had at least two measurements, consistent within the error limits. For example, the corotation radii  distribution for NGC 4123 (see Fig.~\ref{distribution}, top panel)  shows four measurements localized near 60 arcseconds. Note that the 
total error coverage for this galaxy is approximately a quarter of its optical radius (indicated by the brown horizontal line on top of the plot). 
\par
In addition, a significant portion of the collected sample ($\sim$10\%) exhibits several peaks in the distribution with consistent corotation radius positions. Therefore, for the galaxy NGC~864, placed in the middle panel of Fig.~\ref{distribution}, several assumed positions for $R_c$, such as $30$ arcsec, $55$ arcsec and $80$ arcsec can be detected. Note that, for this particular distribution, the total error coverage takes up more than half of the optical radius of this galaxy. The presence of several corotation resonance positions can be attributed to the fact that galaxies may possess multiple spiral modes rotating at different angular speeds~\citep{Rautiainen99, Quillen11}.
Note that the figures for the $R_c$ distributions that have any consistent measurements are available as supplementary files. 
\par
For the remaining part of this dataset, we observe inconsistent\footnote{Note, that using the term ``inconsistency'', we do not mean that incorrect measurements were obtained in the corresponding paper; here we rather point out the discrepancy of the corotation positions estimated by different methods.} corotation radius measurements obtained by different methods. An illustrative example of such disagreement is the distribution of $R_c$ for the well-recognizable grand-design galaxy M100. The bottom panel of Fig.~\ref{distribution} shows an almost uniformly distributed $R_c$ values. The same result can also be obtained by considering the spirals' angular pattern speed, which changes like the angular velocity of the disc. This phenomenon is known as ``dynamic spirals'', and has been modeled in \citet{Sellwood10}, \citet{Carlberg85}, \citet{Fujii11} and others. In particular, fig.~4 (right panel) from~\citet{Roca-Fabrega13} illustrates that the angular pattern speed is not a fixed value. Then, following the corotation radius definition, its location can be found  almost at every galactocentric radius in the disc, as the bottom panel of Fig.~\ref{distribution} demonstrates.
\par
It is important to note that objects with consistent measurements of $R_c$ have, on average, a relatively small total error coverage. However, there are some objects for which the distribution of their corotation radii do not demonstrate any agreement among measurements, despite their having a relatively small value of $\Sigma_{err}$. Galaxies with a high error coverage fraction certainly have inconsistent corotation radius positions. Thus, the magnitude of the total error coverage does not always indicate whether $R_c$ values of the same object are consistent or not. Therefore, we need to look at another quantity that measures the degree of consistency in the measurements. In the next Section, we will examine the distribution of $\Sigma_{err}$ values and another measure of disagreement for the entire sample of galaxies.

\section{Analysis of measurement consistency}
\label{sect:discussion}

The analysis of the corotation radii distributions shows that only about 25\% of the considered galaxies have consistent measurements. The other objects reveal completely disagreement values, either with rather large coverage error or with a small one. In order to differentiate between these distributions, we considered another measure of discrepancy in the locations of corotation radii, which we have named \textit{consistency}. This is calculated as the ratio of the average difference between measurements and the average variance:
\begin{equation}
 \text{\textit{consistency}} = \dfrac{\sqrt{\smashoperator[r]{\sum_{i=1}^{n-1}} \sum_{j=i+1}^{n} (r_i - r_j)^2}}{\sqrt{\smashoperator[r]{\sum_{i=1}^{n}} \sigma_i^2}},
\label{eq:consistence}
\end{equation} 
where $r_i$, $r_j$ are corotation radius values and $\sigma_i$ is a magnitude of error. The larger this value is, the more discrepancy of corotation radius positions is in the galactic disc. This parameter was obtained based on the so-called dimensionless separation of means, as described in \citet{Ashman1994}. It has been used for astronomical data before \citep{Muratov2010,Gusev2020}, but we have modified the original formula to extend its application to more than two measurements. Note that this parameter measures consistency qualitatively, not quantitatively. In other words, it does not have a specific value that separates consistent corotation positions from inconsistent ones.  
\par
\begin{figure}[ht]
   \centering
  \includegraphics[width=15cm, angle=0]{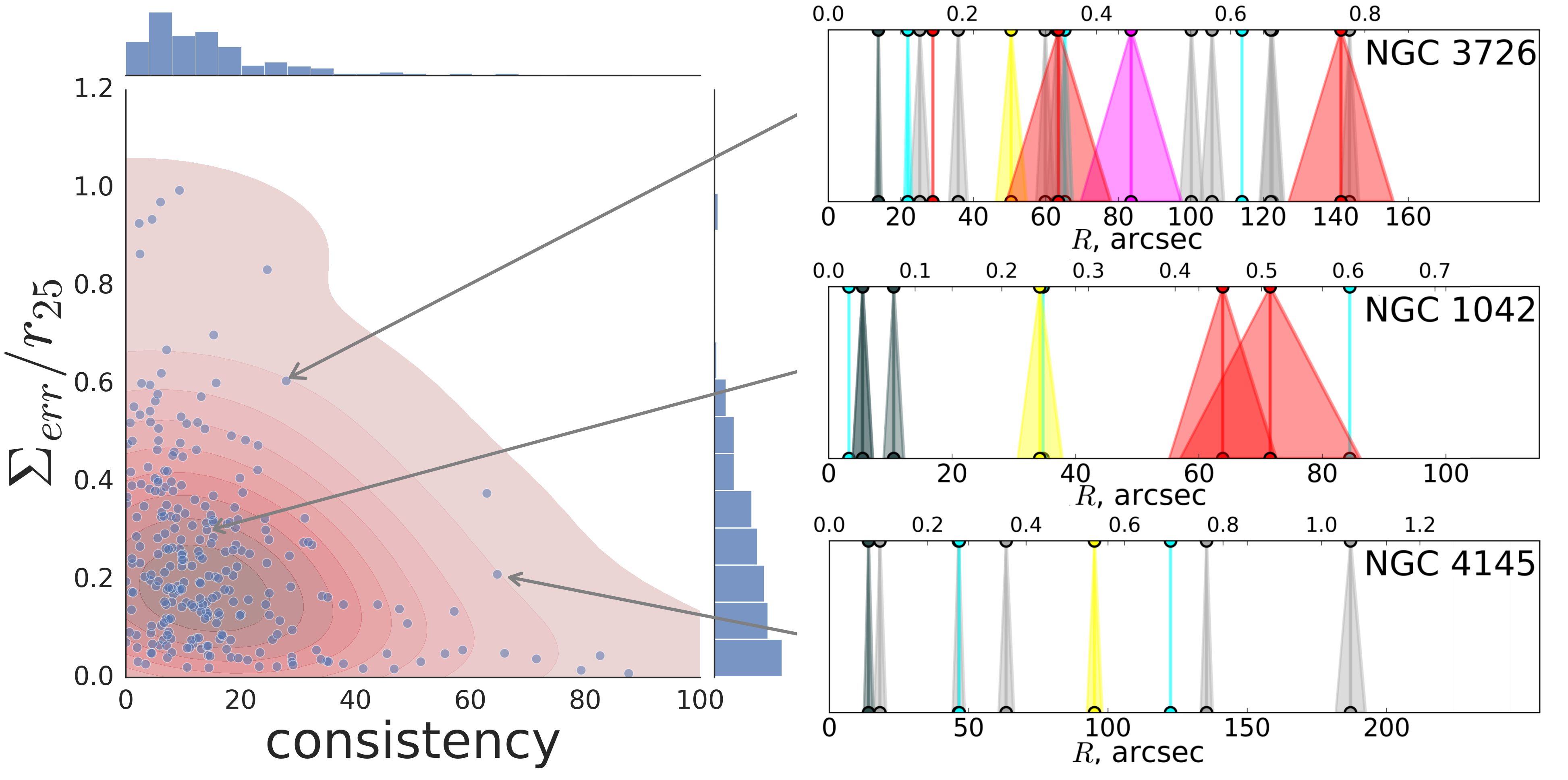}
  
   \caption{The diagram in the left panel illustrates the dependence between error coverage fraction and the measure of \textit{consistency}. Each point corresponds to a galaxy with at least two measurements. The histograms of the corresponding values are presented on the right and at the top. On the right panel, there are examples of $R_c$ distributions (similar to Fig.~\ref{distribution}) for the objects indicated by the arrows. } 
\label{consistency}

   \end{figure}

Figure~\ref{consistency} demonstrates the distributions of the total coverage error quantity and the measure of \textit{consistency} for each object in our dataset. This figure visually separates the sample into different groups. Galaxies with a localised corotation radius are located in the bottom left position on this plot. The distributions with an intermediate $\Sigma_{err}/r_{25}$ value are expected to have multiple $R_c$ values (see the distribution for NGC~1042 in the right panel). Other points indicate the objects for which the estimated corotation radius values from different methods do not agree well. The top and bottom distributions in the right panel of Fig.~\ref{consistency}  provide an illustrative example of how measurements can be inconsistent with large and small coverage errors, respectively. 

\begin{figure}[ht]
   \centering
  \includegraphics[width=8cm, angle=0]{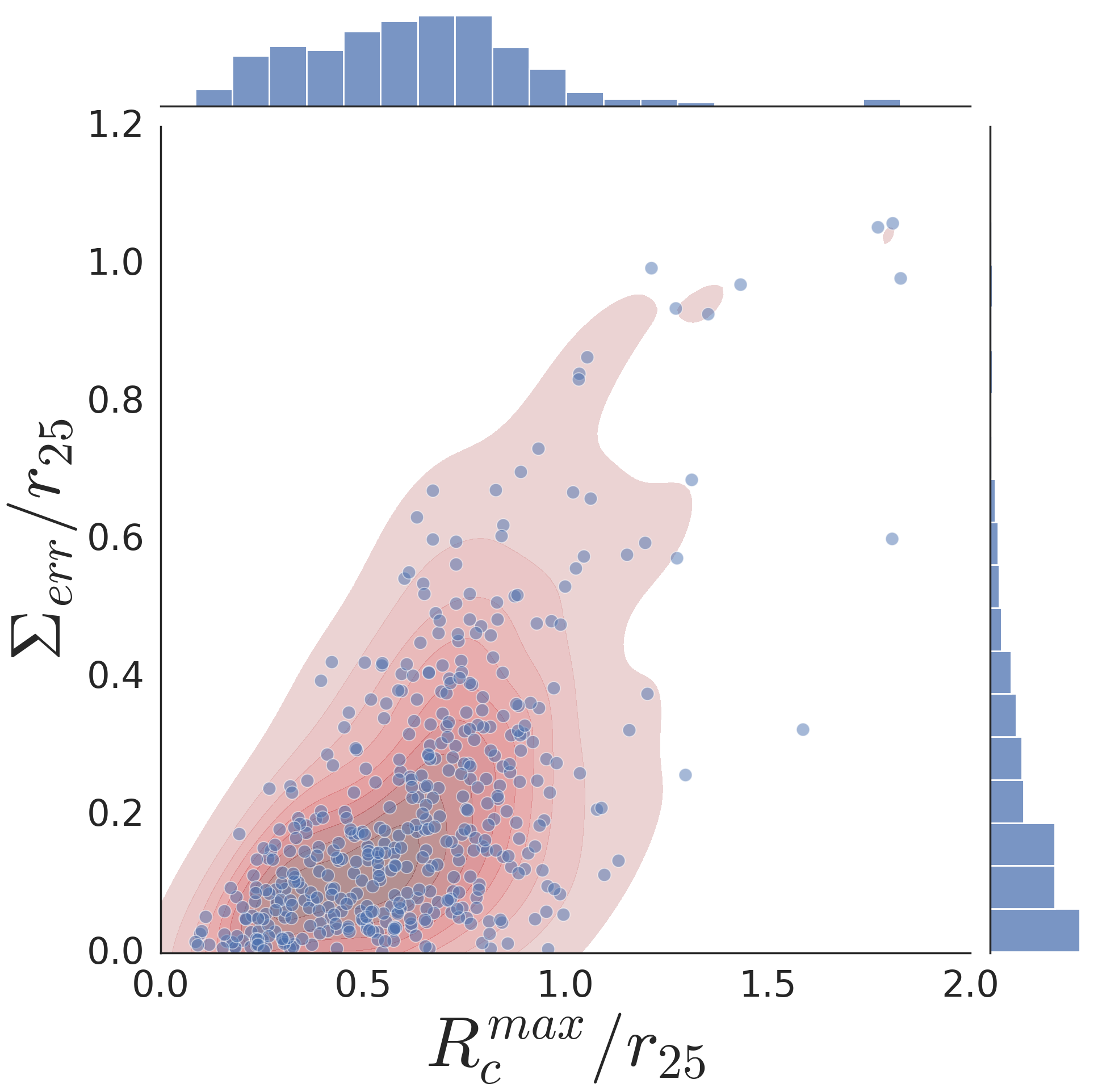}

   \caption{The relationship between the coverage of error in measuring the corotation radius and the maximum of measured $R_c$ value, both quantities normalized by the optical radius $r_{25}$. Above and to the right of the graph are density distributions of the corresponding quantities.} 

\label{cover_cr}
   \end{figure}

Figure~\ref{cover_cr} shows the relation between $\Sigma_{err}/r_{25}$ and the distance from the galactic centre where the maximum corotation radius was measured. This means that the farther from the galactic centre the $R_c$ was measured, the greater the fraction of the galaxy's disc was taken by assumed corotation radii. It is important to point out that, according to this figure, the corotation resonance may have a position beyond the optical radius. This is possible because $R_{c}^{max}$ was defined as the maximum value within the error limits. Besides, the galaxies with barely visible spiral arms extending beyond the $r_{25}$ are not a rare phenomenon \citep[see][]{Mosenkov24}. The observed dependence on Fig.~\ref{cover_cr} has several interpretations. Firstly, this trend does not contradict to the suggestion of the existence of one or more corotation resonances for each galaxy. In this case the coverage error is denoted by the number of measurements and their typical error magnitudes. Hence, the further away from the centre the last corotation measurement is, the more resonances a certain galaxy can have, and the greater magnitude of the error coverage can be calculated. Secondly, some objects may not have localised corotation positions. This is possible for galaxies with dynamic spirals, where the corotation radius can literally be at any point of the disc. In this case $\Sigma_{err}$ would actually increase, due to the expanded area of the disc covered by the measurements. The last option is to assume that most of the methods used or estimates of the errors are incorrect. This interpretation is the least likely, since in this case, Fig.~\ref{cover_cr} would have a chaotic distribution of points with no clear dependence. However, before drawing any considerable conclusions, we must investigate other factors that might be responsible for this discrepancy.\par

\begin{figure}[ht]
   \centering
  \includegraphics[width=12cm, angle=0]{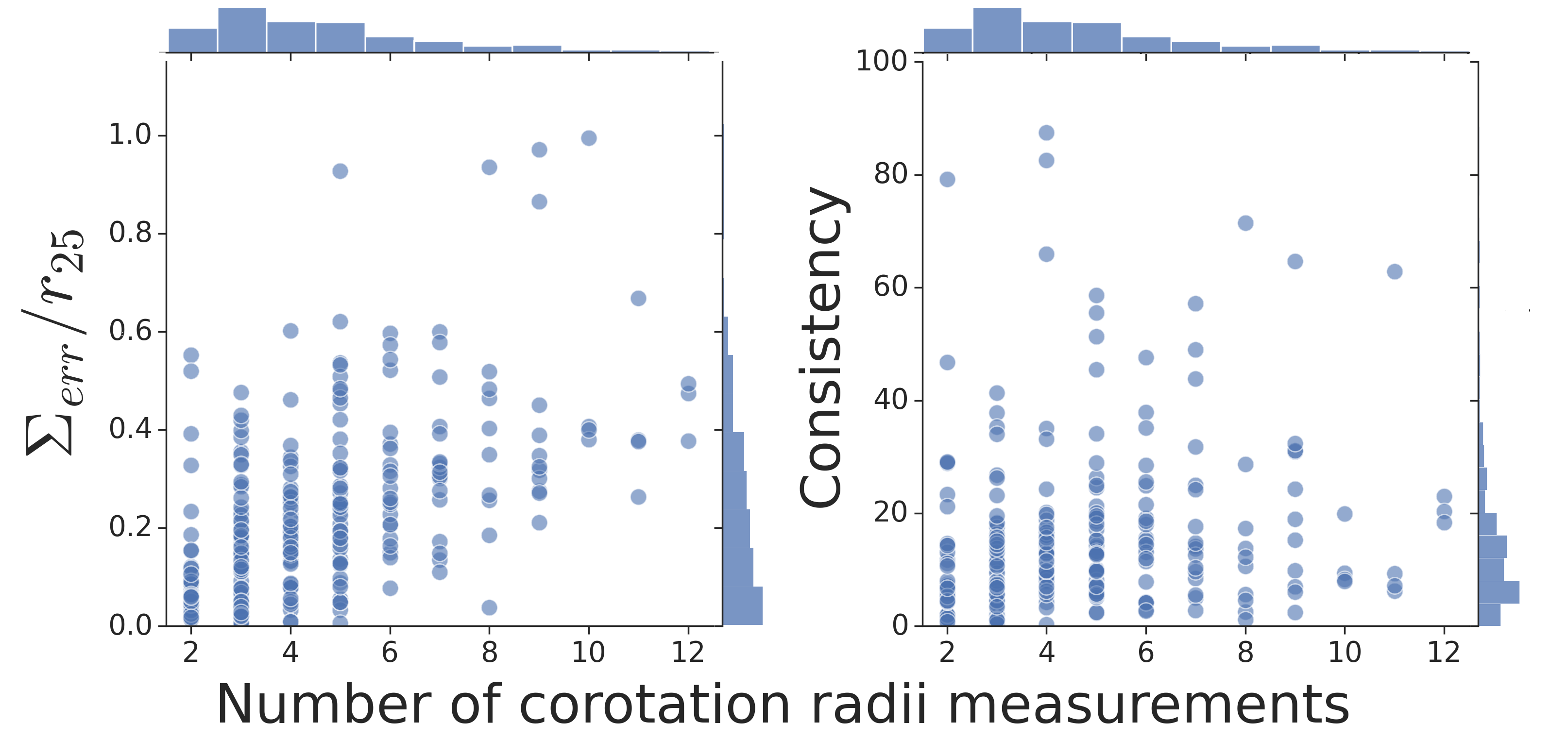}

   \caption{The dependence between the measure of \textit{consistency} (right panel), the coverage of error (left panel) and the number of $R_c$ measurements. Density distributions of the corresponding quantities are shown above and to the right of the graph.} 

\label{cr_num}
   \end{figure}

   Firstly, the measure of \textit{consistency} and the magnitude of total error coverage can increase with the number of measurements found for each galaxy. Figure~\ref{cr_num} does not explicitly show a dependence. However, according to the left panel of this figure, the error coverage magnitude increases, but not steadily, and objects with over seven corotation radius measurements have a significant fraction of the visible galactic disc covered by $\Sigma_{err}$. Conversely, the right panel of the same figure reveals that the majority of objects with inconsistent measurements possess fewer corotation radius values. Additionally, the measure of \textit{consistency}, on average, is slightly lower for numerous measurements.
   
\begin{figure}[ht]
   \centering
  \includegraphics[width=15cm, angle=0]{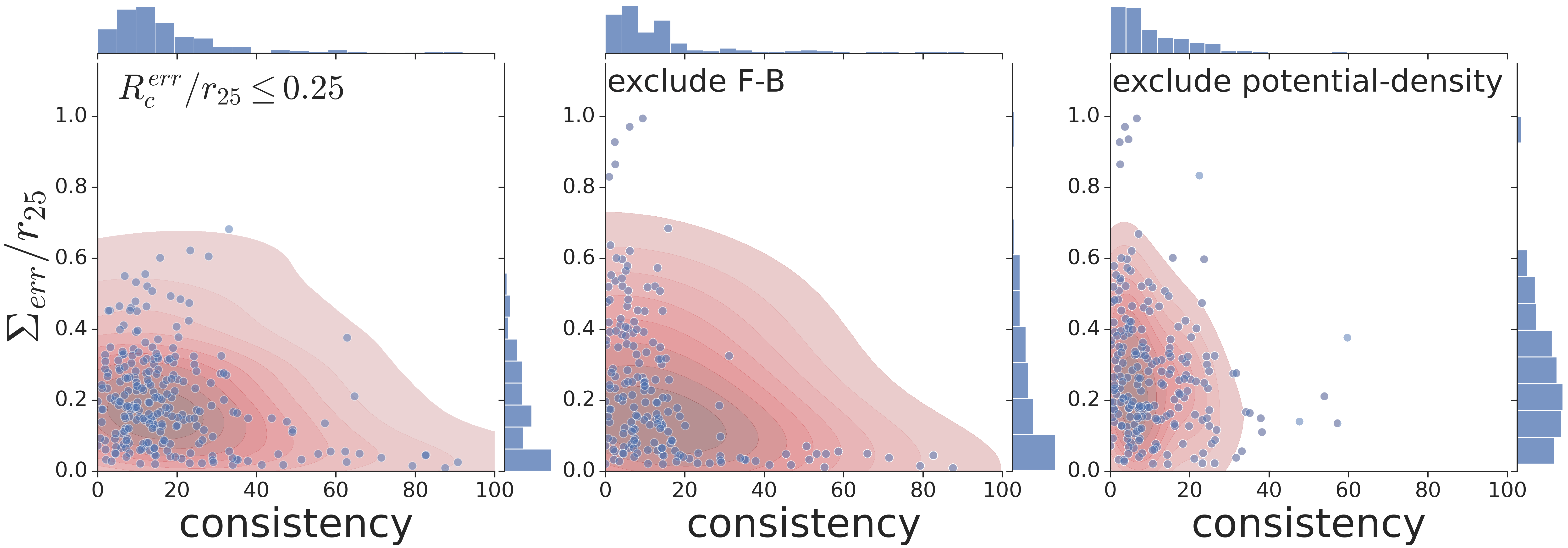}
  
   \caption{Every panel is the same as in Fig.~\ref{consistency}, but excluding corotation radius values which uncertainties exceed a quarter of $r_{25}$ (left) and estimated by \textit{F-B} (middle) and \textit{potential-density} (right) methods. } 
   \label{exclusion}
\end{figure}

Secondly, a significant contribution to the magnitude of $\Sigma_{err}$ can be made by measurements with large errors. Analysis shows that, for most cases, the corotation resonance locations are estimated with errors that are not exceeding a quarter of the optical radius of the galaxy, and only a few measurements have extremely large errors (more than $0.5 r_{25}$). The remaining of corotation radius values with errors less than $0.25 r_{25}$, as expected, lead to a noticeable decrease in the $\Sigma_{err}$ magnitude (Fig.~\ref{exclusion}, left panel). However, even after excluding values with large errors, a significant fraction of galaxies still exhibit a considerable value of total error coverage. Furthermore, the exclusion of values with large errors cannot improve the measure of \textit{consistency} because individual measurements can be completely inconsistent, despite their small errors.
\par
Other possible reasons for the discrepancy of $R_c$ values may be related to the assumptions and accuracy of certain methods, the reliability of the determined values, and the estimation of errors. Methods that primally measure several corotation radii locations for one galaxy apparently have an influence on the revealed tendency. To investigate this aspect, we removed measurements estimated by such methods one at a time. The exclusion of measurements obtained by \textit{F-B} method (the middle panel in Fig.~\ref{exclusion}) demonstrates a relatively small number of objects with intermediate values of $\Sigma_{err}$ fraction and \textit{consistency}. As discussed earlier, this method can determine not only corotation resonance positions. So, these measurements do not mostly intersect, and the median value of their coverage error is about $0.2 r_{25}$ (see Fig.~\ref{method_err_violin}). Consequently, the measure of \textit{consistency} and $\Sigma_{err}$ fraction do not need to have small or large magnitudes in those galaxies where the \textit{F-B} method has been applied. The next method, whose accuracy was examined, is the \textit{potential-density} method. Its multiple measurements for a single galaxy indicate the presence of several spiral modes. Also, those values have almost zero errors. This explains the disappearance of points with large measure of \textit{consistency} (Fig.~\ref{exclusion}, right panel). It is important to note that these extremely large \textit{consistency} values were caused by near-zero measurement errors rather than by their distance from each other. Despite the exclusion of \textit{potential-density} method measurements, the measure of \textit{consistency} still have significant magnitudes. In addition, noticeable effects on $R_c$ distributions could be caused by those methods whose measurements have large errors, such as \textit{T-W} or \textit{offset} (see Fig.~\ref{method_err_violin}). However, the exclusion of these methods` measurements showed minimal differences compared to Fig.~\ref{consistency}.
Thus, this analysis shows that the inconsistency of corotation radius values is more related to the fact that, for some objects, these measurements of $R_c$ can be distributed evenly throughout the disc (the case of transient spirals), rather than due to the reliability of any specific method.

\begin{figure}[ht]
   \centering
  \includegraphics[width=12cm, angle=0]{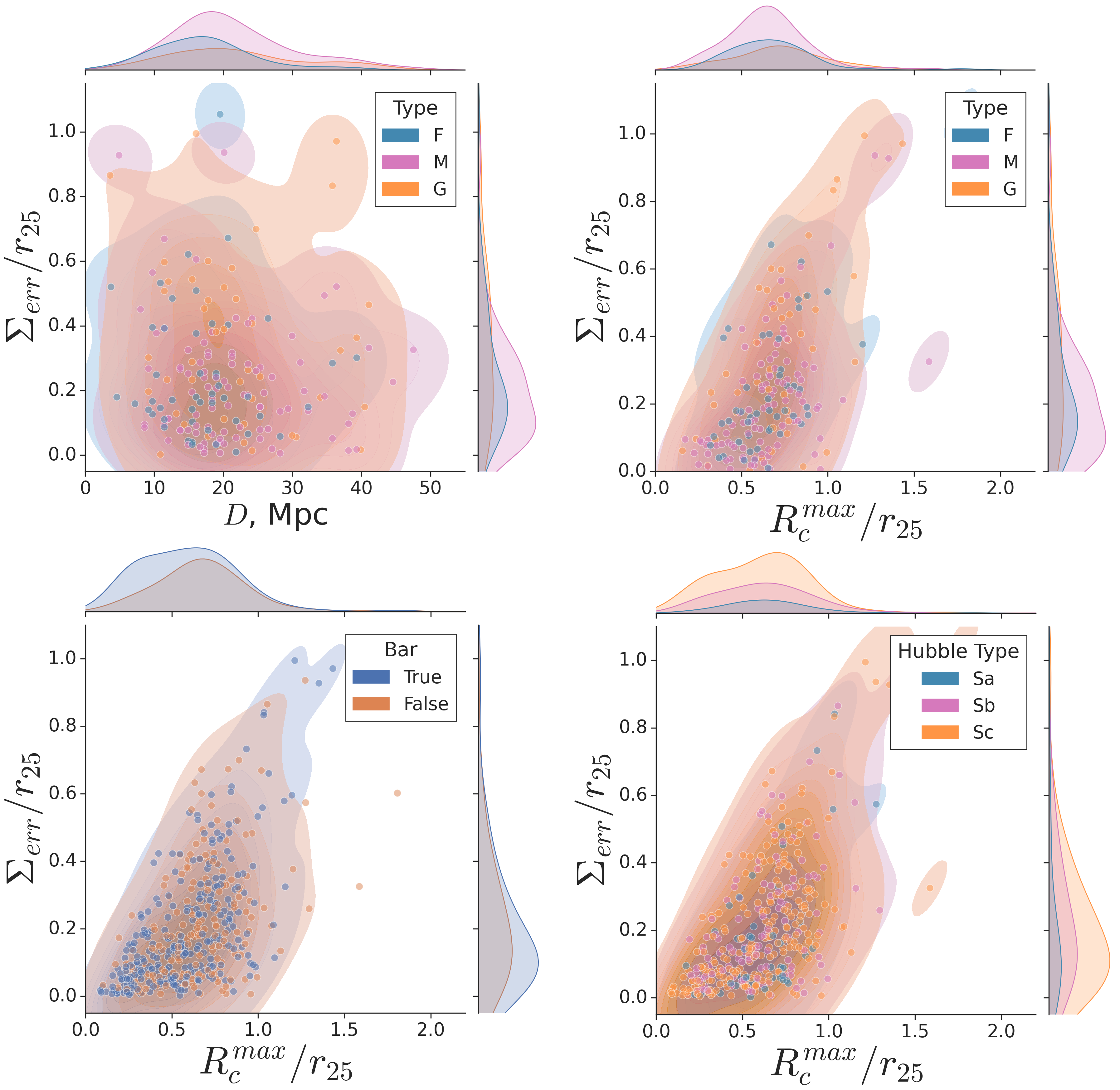}
  
   \caption{Dependencies of the coverage fraction error in estimating the corotation radius (ordinate axis) from the farthest value $R_{c}$ (top right and two bottom) and from the distance to objects (top left). The colour on the top two images indicates the type of spirals: F~--- flocculent, M~--- multi-armed, and G~--- grand design \citep[determined in][]{types}. The colour of the points on the lower right figure is determined by the Hubble type, and on the left, it indicates the presence of a bar. Density distributions of the corresponding quantities are shown above and to the right of the graph. Distances, Hubble classification and presence of a bar data were obtained from~\citet{Hyperleda}. Note that the number of points on the diagrams depends on the number of objects for which the morphological type and spiral pattern type are known.} 
   \label{cover_cr_analysis}
   \end{figure}

In addition, we investigated whether there is any selection effect due to the distance of the galaxies. Figure~\ref{cover_cr_analysis} (top left) shows that the total coverage of errors is not dependent on this parameter. Furthermore, the galaxies with distinct spiral structure (basically, grand design or multi-armed) are believed to match the spiral density wave theory more closely than flocculent ones \citep{LinShu1967,Thomasson1990,Dobbs2007}. However, as we can see from the top right panel, galaxies belonging to all of these spiral types can exhibit both a small and large fraction of $\Sigma_{err}$. Furthermore, we found that there is no division in \textit{consistency} magnitude between grand design, multi-armed and flocculent galaxies. This implies that galaxies with prominent spirals can have inconsistent measurements that may indicate their multiple pattern or dynamic spirals nature. 

\par
Classical density wave theory only considered tightly wound spiral arms, so the magnitude of $\Sigma_{err}$ can be related to the pitch angle value. To investigate this point, we consider the distribution of Hubble type galaxies. However, when posing the question in this way, the presence of a bar in a galaxy may not be taken into account. Therefore, in the legend of Fig.~\ref{cover_cr_analysis} (bottom right), for example, the types SBa and Sa are considered equivalent. Nevertheless, there is no sharp division in total error coverage magnitude between galaxies with open spirals and tight-wound ones. The bottom left panel of Fig.~\ref{cover_cr_analysis} does not demonstrate any dependence on whether a galaxy has a bar or not. Note that the density distributions (blue) above the figure have a double peak, and the left one is related to objects that have measured the corotation radius of their bar. To find out whether a bar is present in a galaxy or not, we use morphological type and the presence of bar data from HyperLeda~\citep{Hyperleda}. Additionally, if possible, we exclude weakly barred galaxies from the sample of barred galaxies. 

While the scenario of transient spiral arms may be more suitable for explaining our results, all of the cases considered above are theoretically possible. Moreover, it is likely that some of them actually occur in practice. The graph on the right of Fig.~\ref{consistency} contains a variety of objects with different nature of spiral structures. Unfortunately, it is not yet clear which mechanism predominates in most spiral galaxies. Therefore,  we need to examine each individual galaxy based on the collected dataset.

\section{Winding time of spiral arms in galaxies}
\label{sect:wind_time}

The collected data is indeed will be useful for clarifying various aspects of the spirals' nature. In this Section, we will demonstrate an example of how this data can be used to answer a specific question related to their lifetime. As we discussed earlier, the spiral structure of disc galaxies may consist of several spiral modes~\citep{transient}. Each of these spiral substructures rotates with their own pattern speed, and therefore they have different corotation resonances. According to the~\citet{Quillen11} and~\citet{Minchev12} papers, these substructures may interact with each other in a special way: the $R_c$ of the slower rotating pattern coincides with another resonance of the faster rotating ones. Spiral density waves moving at different angular velocities often intersect at the outer Lindblad (OLR) or ultraharmonic resonances of the internal spiral pattern~\citep{Rautiainen99}. Their positions are defined as distances at which the pattern speed is equal to $\Omega+\kappa/2$ and $\Omega+\kappa/4$, where $\kappa$ is the epicyclic frequency and $\Omega$ is the angular velocity of the disc rotation. Will such a spiral structure wind up and destroy its arrangement in such a short period of time, like material spirals do~\citep{Oort1962}? To investigate this issue, we can directly calculate the winding time value of the whole spiral pattern using the pattern speed of each spiral mode~\citep{Meidt08}:
\begin{equation}
   \label{eq:winding_time}
   \tau_{\text{wind}}= 2 \pi / (\Omega_p^{\text{inner}} - \Omega_p^{\text{outer}}),
\end{equation}
 where $\Omega_p^{\text{inner}}$, $\Omega_p^{\text{outer}}$ are the angular velocities of the inner and outer spiral structure, correspondingly. Technically, this quantity means the time after which the inner substructure will overtake the outer ones by the whole revolution around the galactic centre.
\par
To investigate this aspect, we focus on five objects from our dataset that could have multiple spiral modes: IC~342, M~58, M~74, NGC~3583 and NGC~5371. The $R_c$ distributions of these galaxies have consistent measurements and demonstrate the evident existence of several corotation resonances. In addition, the rotation velocity profiles for these objects are also publicly available. The collected sample is not a comprehensive list of all objects with described structures; the limited availability of rotation curves in literature limits the size of our sample.

\begin{figure}[!ht]
   \centering
  \includegraphics[width=12cm, angle=0]{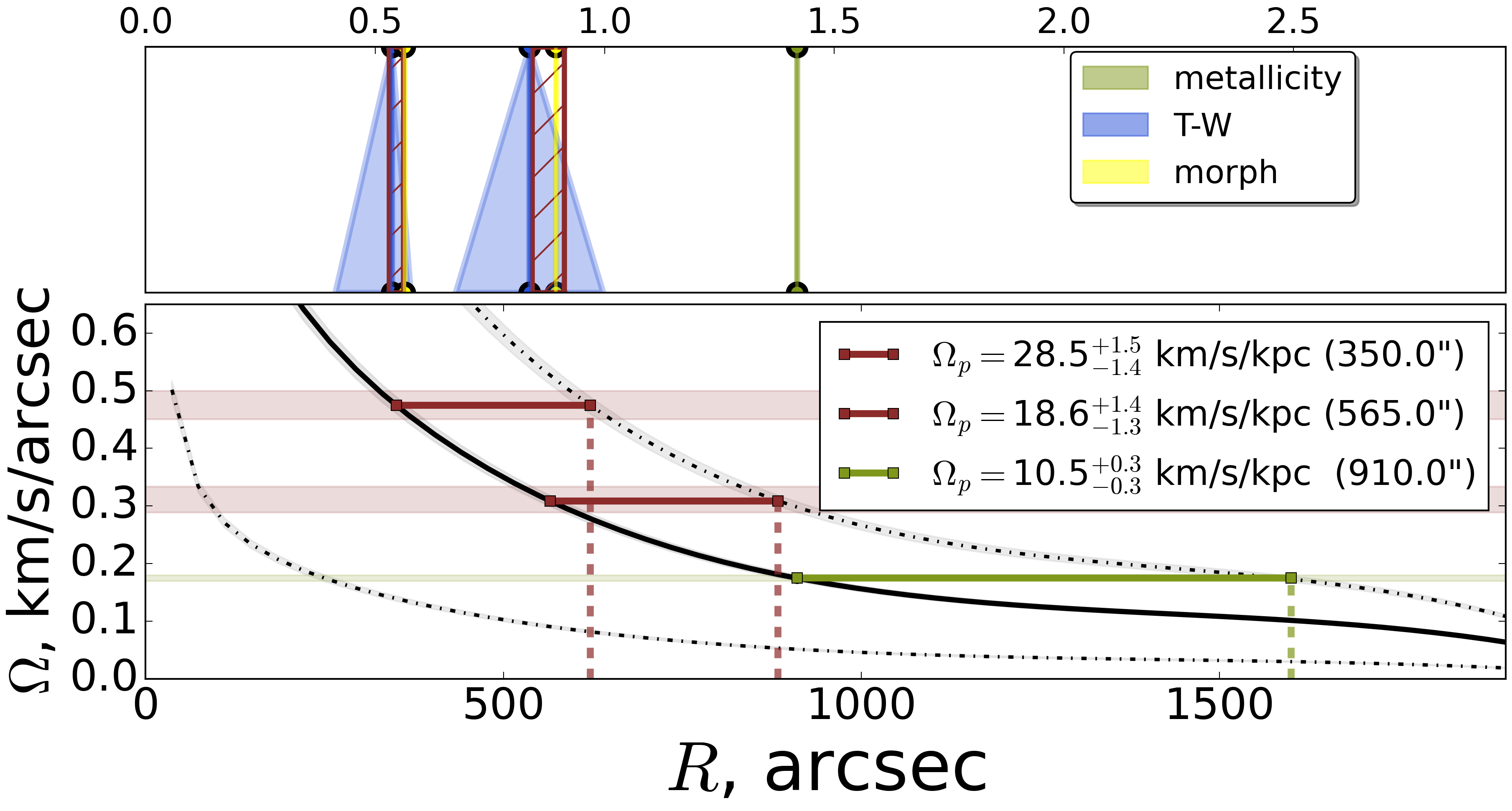}
  
   \caption{\textit{At the top}: vertical lines show the $R_c$ errors (shaded areas). Their colour indicates the method (see the legend). The upper scale determines corotation radius positions in units of the optical radius. The intersection of the measurements of the radius of the corotation is highlighted by a maroon hatched area.\\
\textit{At the bottom: }a solid black line shows the angular velocity curve of the disc with an error (gray shaded area). To the left and right of it, dashed lines define the areas of inner and outer Lindblad resonances. The shaded maroon horizontal areas correspond to the errors of the angular velocity of the pattern, calculated on the basis of the radius of rotation found at the intersection of measurements of this value from the upper figure. The horizontal line connects the positions of the corotation radius and the OLR for the angular pattern speed, the value of which is presented in the legend. Note, that x-axis of top and bottom plots has the same scale.} 
   \label{coupling}
\end{figure}
\par
Subsequently, we obtained rotation curve data $v_c(R)$ for these five objects from the following sources:~\citealp{Crosthwaite2000} (IC~342),~\citealp{Lang2020} (M~58),~\citealp{Walter2008} (M~74),~\citealp{Noordermeer2005} (NGC~3583),~\citealp{Sanders1996} (NGC~5371). Taking IC~342 as an example, we conducted a detailed analysis, presented  in Fig.~\ref{coupling}. In the top panel, we marked the supposed corotation radius positions, with the maroon hatched areas indicating $R_c$ values at the intersection of several measurements. Using the obtained angular velocity profile $\Omega(R) = v_c(R)/R$, we determined the pattern speeds of each mode, as shown by the vertical lines and the legend in the bottom figure. Additionally, we plotted the $\Omega\pm\kappa/2$ curve indicated by dashed line. As we can see,  considering the uncertainties in the locations of resonances and the pattern speed, the OLR positions of the internal structures are in good agreement with the corotation radii of the external ones. This provides a strong evidence for the existence of multiple spiral modes. Note that two corotation positions estimated by the \textit{T-W} method were taken from \citet{Meidt09}, which also detected a similar feature in this object. In this paper, we have confirmed these $R_c$ locations with independent measurements obtained from \citet{Roberts75,Elmegreen(1992)} and showed the presence of the external structure rotating with 10~km/s/kpc. Analogous phenomena have been observed in other selected galaxies as well.

\par
\begin{figure}[ht]
   \centering
  \includegraphics[width=11cm, angle=0]{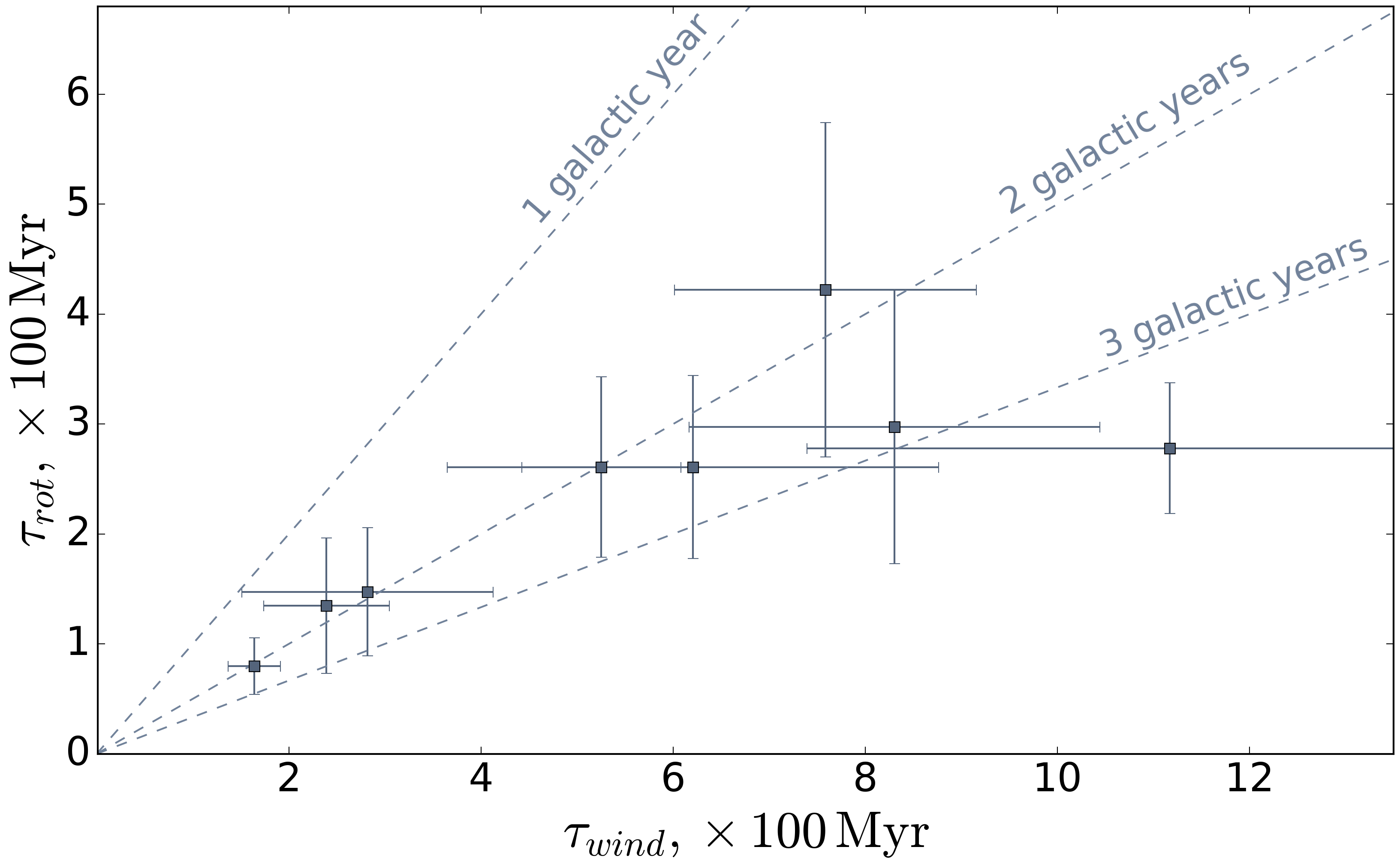}
  
   \caption{The figure demonstrates a comparison between the winding (X-axis) and rotating (Y-axis) times. Horizontal error bars evaluated from the uncertainties of angular velocity values. Note, that the number of points in the graph represents the number of pairs of adjacent substructures, rather than the number of galaxies.}  
   \label{winding_time}
\end{figure}

Furthermore, by considering formula~(\ref{eq:winding_time}), we calculated the winding time and compared it to the rotation periods $\tau_{rot}$ obtained for each spiral mode in the galaxy. It is worth noting that, according to our analysis, some objects exhibit three spiral modes (as for IC~342), while for others we found only two. The results are presented in Fig. \ref{winding_time}, which shows the galactic rotation period obtained by taking the average of the angular speeds of adjacent the inner and outer spiral modes. The figure illustrates that, for all the galaxies in our sample, the winding up process of the spiral structure takes approximately two or three galactic years. The resulting $\tau_{\text{wind}}$ values are larger than those predicted for material spirals or found for other galaxies~\citet{Merrifield2006}. However, these times are not long enough to consider the spirals as a long-lived structure, and it also means that for such galaxies the ``winding dilemma'' still remains as an unresolved issue. These findings additionally poses a question about the supporting mechanisms that enable the spiral patterns to be unwinding and persist for extended periods of time.

\section{Conclusion}   
\label{sect:conclusion}
The question of the nature of spiral structure in galaxies has been a long-standing problem in astrophysics. Despite the strength of existing theories, it is difficult to find out which of them correspond to real galaxies using observational data. Some research papers provide observational evidence to support~\citep{Tamburro08,Pour-Imani2016,Chandar2017,Peterken19}, or contradict the density wave scenario~\citep{Foyle2011,Shabani2018}. Furthermore, there are some assumptions that all of proposed mechanisms could contribute to the formation of spiral arms in different galaxies. For example, a recent research of~\citet{Chen2023} showed that the spiral arms of two galaxies with similar features could be consistent with both the density wave and the recurrent spiral theories.
\par
One of the key observational tests of the density wave theory relates to the constant pattern speed of the spiral structure, which implies the existence of a localized corotation radius position. A reliable estimate of $R_c$ is possible when measurements obtained by different methods are consistent within the error limits. On the other hand, an almost uniform distribution of corotation radii can be interpreted by the dynamic spirals` theory. However, it is still not clear which mechanism is the dominant one in the real galaxies. To investigate this problem, a dataset of corotation radius measurements of 12 various methods was collected for 547 objects, with 300 of them having more than one value. All gathered data is publicly available on GitHub for anyone who will find it useful for their future research. Note that in our paper, we for the first time performed a meta-analysis on a significant number of studies that estimate the corotation resonance positions by different methods. Although, a similar analysis can be found in other studies, for example, \citet{Vallee20} compared the results of research papers demonstrating the existence or absence of stellar age gradient for 24 galaxies. \citet{Beckman18} also tested the consistency between Tremaine-Weinberg and Font-Beckman methods for a single galaxy. 
\par
We examined the distributions of corotation radii for galaxies with at least two measurements obtained by different methods. About 15\% of these objects have a visually consistent position of $R_c$, while 10\% could have several localized corotation resonances. The remaining galaxies do not show any consistency in their distributions. The discrepancy between measurements for certain objects were found not only among those obtained by different methods \citep{ScaranoLepine(2012)}. As \citet{Williams(2021)} demonstrated, the pattern speed values (and corotation position, respectively) measured by the Tremaine-Weinberg method may vary significantly depending on which component (gaseous or stellar) was used (see fig.~6 from \citealp{Williams(2021)}). Besides, the aforementioned method can produce completely different results, depending on the initial assumption of whether the pattern speed has a constant value or varies arbitrarily with radius \citep{Merrifield2006}. However, the measurements of some methods included in our dataset can differ from each other because their applications imply the existence of multiple corotation resonances~\citep{Buta&Zhang(2009),Font11,F-B14}. 

\par
To investigate the possible reasons for these results (see Fig.~\ref{consistency}), we calculated the total coverage error $\Sigma_{err}$ in estimating the value of $R_c$ and the measure of \textit{consistency} (formula~\ref{eq:consistence}) for each galaxy. The combination of these quantities allows us to divide galaxies with a localized corotation radius position from those with an almost uniform distribution. Our analysis revealed that the coverage error values, on average, increase with the distance to the farthest corotation resonance (Fig.~\ref{cover_cr}). Additionally, the investigation showed that the magnitude of these quantities has a weak correlation with the number of measurements (see Fig.~\ref{cr_num}). Even for galaxies with just a few $R_c$ values, the magnitude of $\Sigma_{err}$ might constitute a significant fraction of the disc.
\par
 We also tested the reliability of the applied methods (Fig.~\ref{exclusion}). The exclusion of the \textit{potential-density} method \citep{Buta&Zhang(2009)} showed that the negligible errors of these measurements did not influence $\Sigma_{err}$, but their multiplicity resulted in a large value for \textit{consistency}.
 As Fig.~\ref{method_err_violin} demonstrate, the various \textit{offset} and \textit{T-W} methods have a relatively large average magnitude of $\Sigma_{err}$. Besides, there are other reasons to suspect these measurements to be ``unreliable''. Firstly, \citet{Borodina2023} demonstrated that the use of a gaseous component in the Tremaine–Weinberg method could lead to incorrect results. Also, most part of gathered \textit{T-W} measurements were obtained based on $H_{\alpha}$ and CO velocity fields \citep{RandWallin(2004),Fathi,Hernandez05}, so they may not be accurate. Regarding the \textit{offset} method, the fact that half of the studies detect a stellar age gradient while the other half questions its existence \citep{Vallee20} does not support the reliability of this method. However, the exclusion of measurements obtained by the aforementioned methods did not lead to a noticeable decrease in the error coverage and \textit{consistency} values. Moreover, we obtained almost the same result by including only those measurements whose errors did not exceed a quarter of the optical radius.
 \par
 According to \citet{LinShu1967,Thomasson1990} the density wave with more than three modes is likely to be unstable, therefore the grand design spiral structure of galaxies is proposed to persist much longer than that of multi-armed or flocculent ones. Another assumption is that the  substructures of flocculent galaxies are formed due to the resonances, stellar feedback, and instabilities in the underlying disc with regular two patterns \citep{Chakrabarti2003,Elmegreen2003}. In addition, \citet{Block1994,Thornley1997,Seigar2003} found examples of multi-armed and flocculent galaxies with an underlying grand design structure traceable by K-band, although more recent observations showed that most of flocculent galaxies do not have such features \citep{Elmegreen2011}. Many studies were dedicated to investigate the distinctive properties of galaxies exhibiting spirals of different types \citep[for example,][]{Ann2013,Bittner2017, Willett2013, Hart2017,Savchenko20}. However, it is still unclear whether such galaxies obey to different scenarios of spiral formation. Our analysis (see Fig.~\ref{cover_cr_analysis}) has revealed that there is no clear division between grand design, multi-armed, and flocculent galaxies in terms of both the total coverage error and measure of \textit{consistency}. This implies that, assuming the reliability of all the methods and their measurements, galaxies of all spiral types could have one or more or almost evenly distributed corotation positions. Thus, according to the data in the graph, the type of spiral structure does not necessarily determine its nature. In addition, we did not find any dependence on Hubble classification (pitch angles). Also, we did not reveal any relation with the distance of galaxies and the presence of a bar (see Fig.~\ref{cover_cr_analysis}).

 \par
 Although the obtained results are consistent with a picture when significant part of galaxies contain spirals of transient origin \citep[there some evidence for this, for example,][]{Pringle2019,Masters2019}, we can not exclude a fact that a different combination of above-mentioned reasons could be realised for every object. This multi-level problem is difficult to solve in practice. That is why it is necessary to investigate each galaxy individually (for example, as in \citealp{Beckman18}).
\par 

In addition to clarifying the discrepancy in $R_c$ measurements for a collected sample of objects, it was calculated a winding time value (formula~\ref{eq:winding_time}) for several galaxies in our dataset. According to Fig.~\ref{winding_time}, we found that the spiral structures of these objects are winding up on a timescale equal to several orbital periods. Our results are consistent with the picture shown in \citet{Merrifield2006,Meidt09}. There is a tendency for recent models and observations to move away from the stationary density wave scenario \citep{Sellwood2011}. For example, \citet{Masters2019} examined a  large sample of nearby galaxies and found no correlation between spiral arm winding tightness and bulge size predicted by density wave theory. Besides, recent simulations have also provided evidence for transient spiral structures in external galaxies, as well as our own Milky Way \citep{D'Onghia2013,Hunt2018,Pettitt2018,Sellwood2022,Asano2024,Funakoshi2024}. However, convincing evidence in favour of density waves in most of the galaxies also continues to emerge \citep[for example,][]{Davis2015,Yu2018,Yu2019,Peterken19}. These contradictory findings highlight the complexity of the spiral nature problem and the necessity for development of a new approach to solve it.
\par

 Despite the fact that the compiled dataset for galaxies consists of quite heterogeneous measurements taken from various sources, it will still be valuable for further investigations. Firstly, we can determine the angular speeds of galactic spiral structures by analysing the corotation radius positions and the rotation curves of the corresponding galaxies. This provides an opportunity to examine galaxies with multiple spiral patterns and investigate the coupling effect in more detail. Secondly, the obtained angular speeds of the spiral patterns can be used to explore the dependence between bar and spiral structures. Additionally, by obtaining reliable rotation radius values for a statistically significant number of galaxies, we will be able to investigate the relationships between the angular pattern speed and other galactic properties. Thus, in our future paper, we will examine several galaxies in detail using collected dataset. Besides, we will implement and apply some methods (such as in~\citealp{Marchuk2024}) for corotation radii measuring, which could allow us to solve spiral's nature case for certain galaxies. 
\normalem
\begin{acknowledgements}

V.S. is acknowledged support from ``BASIS'' Foundation for the Development of Theoretical Physics and Mathematics (grant No. 23-2-2-6-1).
\par
We would like to thank the anonymous reviewer for his/her thorough review
and for the valuable comments which have helped to improve the quality of this article.

\end{acknowledgements}

\bibliographystyle{raa}
\bibliography{mc2024-0071}

\appendix
\section{Methods}
\label{sect:methods}

In this Section a brief description for each method mentioned in this paper is presented. The conditional names for each of these are indicated in parentheses.

\subsection{Age gradient method (\textit{offset})}
One of the generally accepted ideas about the mechanism of maintaining a spiral structure in a galaxy is based on the theory of quasi-stationary density waves \citep{Lin&Shu}. It assumes rigidly rotating spirals in a differentially rotating galactic disc. The corotation radius, as it follows from definition, divides the galaxy into two parts: an internal and an external. In the central part, the angular velocity of the disc is greater than the pattern speed. This means that, assuming trailing spiral rotation, this means that the gravitational shock wave induced by the arms will trigger star formation on the inner side of the spirals. In the external part, vice versa, the disc rotates  more slowly than the spiral patterns, which is why newborn stars form on the outer side of the pattern. Thus, an azimuthal age gradient across the spiral arm should take place in the galactic disc.
\par
~\cite{Oey} calculated spatial isochrons of the spirals, the intersection of which indicated $R_c$ position (see fig.~7, fig.~8 from~\citealp{Oey}). There are a number of papers investigating the profile of the angular offset ($\Delta\phi$) between young massive star clusters and star-forming regions (\citealt{Sakhibov21},~\citealt{Tamburro08},~\citealt{Egusa09}). The distance on which the $\Delta\phi$ changes sigh is associated with the corotation resonance position. Investigations of azimuthal age gradients can also be performed by comparing spiral arms form between multiwave images of the same galaxy~\citep{Abdeen20}. 
\par
\subsection{Tremaine-Weinberg method (\textit{T-W})}
This method is the only one which estimate the angular pattern speed using observational data~\citep{T-W}. Then, using the angular speed profile of the disc, it is possible to detect a corotation radius position.  
\par
The method can be applied only when the following three conditions are satisfied:
\begin{itemize}
    \item The galaxy disc is flat;
    \item The disc contains a single, well-defined pattern speed;
    \item The tracer obeys the continuity equation.
\end{itemize}
To find the angular velocity of the spiral pattern, the following expression should be applied:
\begin{equation}
   \Omega_p \sin i = \dfrac{\langle v \rangle}{ \langle x \rangle}, \nonumber
\end{equation}
where $\langle v \rangle$ and $\langle x \rangle$ are the intensity-weighted velocity and the position along a slit, respectively, and $i$ is the galaxy inclination. The slits for each object are drawn parallel to its large semi-axis (see fig.~14 in~\citealp{Williams(2021)}). Velocity data along the slit are often extracted from molecular hydrogen (\citealp{Williams(2021)}, \citealp{RandWallin(2004)}, \citealp{Meidt09}) and $\text{H}_{\alpha}$ velocity maps (\citealp{Fathi}, \citealp{Hernandez05}, \citealp{Toonen08}). In addition, some authors use the stellar component velocity map~\citep{Cuomo20}.
\par

\subsection{Metallicity gradient method (\textit{metallicity})}
The metallicity distribution profile in galaxies is characterized by a monotonous decrease in the content of heavy elements from the central regions to the peripheral ones~\citep{Vila-Costas92}. As a rule, such profiles can have breaks or changes of slope caused by the presence of corotation resonance. The corotation radius is considered to divide the galactic disc into two isolated parts that almost do not exchange gas and, as a result, evolve independently. It is important to note that the same effect was found in the Milky Way disc~\citep{Lepine11} and in the hydrodynamical simulations~\citep{Lepine01}. \citet{ScaranoLepine(2012)} investigated the metallicity distribution profiles and determined $R_c$ positions for 27 galaxies. To check the correctness of their measurements, the authors compared obtained results with corotation radius values taken from the literature.
\subsection{Potential-density method (\textit{potential-density})} 
\par
\citet{Zhang96,Zhang98} have shown that the disc matter and the density wave can exchange energy and momentum. This occurs as a result of a local gravitational instability or at the potential minimum of the pattern. Consequently, a shift between the potential distribution and the density wave profiles should be detected. Inside the corotation radius, the spiral wave rotates more slowly, which is why the disc particles lose their angular momentum. A positive sign of the phase shift defined as the azimuthal offset between the potential and the spiral in the direction of the galaxy rotation means that the spiral density wave gains the angular momentum. A negative sign, on the contrary, indicates a loss of momentum. Therefore, the corotation resonance position is where the phase shift changes from positive to negative. 
\par
One of the first applications of this method was carried out in~\citet{Zhang07}. The authors used NIR images ($1.65$ $\mu$m) to obtain surface density and potential distribution maps. The value of the potential was determined by the Poisson equation, and the phase shift magnitude was calculated based on equation (3) from~\citet{Zhang07}. As a result, the phase shift profiles were obtained for 153 objects (see fig.~2 in~\citealp{Buta&Zhang(2009)}). A majority of these profiles have multiple changes of sign from positive to negative values. This fact is evidence of the existing of several spiral modes rotating at different angular speeds. Additionally, these profiles show the locations where the phase shift switches from negative to positive values. It is hypothesized that spiral mode coupling occurs at these positions.
\par
\subsection{Font-Beckman method (\textit{F-B})}
The main concept of this method based on radial gas flows presents near resonances, including corotation ones~\citep{Kalnajs78}, which causes a change in the sign of the radial component of the velocity. The corotation radius estimation method, based on these assumptions, was first applied in~\citet{Sakhibov87,Sakhibov89,Canzian93}. Also, a similar method has been proposed in~\citet{Lyakhovich97}. A general part of corotation radius values measured by similar method in our sample are taken from~\citet{Font11,F-B14,Font_et_al.(2014)}. This method is based on the search for the positions of the so-called ``zero'' non-circular velocities, which occur near resonances. To use the \textit{F-B} method, residual velocity maps are required instead of radial velocity maps, as in \citet{Lyakhovich97}. These residual velocity maps strongly depend on the choice of the inclination angle and the positional angle of the galaxy. It is important to note that this method allows to estimate not only corotation resonances.  
\subsection{Simulation of observable galaxies (\textit{model})}
Another approach to estimating the angular velocity is to simulate individual galaxies. The object model can be constructed based on N-body simulations \citep{Salo2000} and hydrodynamic simulations in an external potential (\citealp{Lindblad96},~\citealp{Rozas08}) both separately and jointly \citep{RautiainenLaurikainen(2008)}. The first step in these simulations is to construct the potential based on the stellar mass distribution obtained from photometric images. The stellar mass of a galaxy generally consists of old population material, so near-infrared (NIR) data is mostly used to determine this value. It is important to mention that some works, like \citet{Kranz03}, include a dark matter potential derived from velocity curve data. Next, the galaxy is modelled in the obtained potential with a fixed angular velocity of the spiral structure. Then, model parameters are varied so that the morphological features of the synthetic and observed galaxies have a similar appearance.
\subsection{Morphological method (\textit{morph})}
The presence of the corotation resonance in the galactic disc promotes the appearance of certain morphological features. As assumed in~\citet{Roberts75}, corotation radius influence the extensions of the disc and the spiral structure as well as the existence of ionized hydrogen regions. In another series of works (\citealp{Elmegreen(1992)},~\citealp{ElmegreenElmegreen(1995)}), additional morphological features have been identified in order to measure the $R_c$ value. One of them is based on the fact that the inner and outer Lindblad resonances have their clear location in the galaxy. The positions of these resonances are defined as the distances at which the angular velocities of the spirals are equal to $\Omega-\kappa/2$ and $\Omega+\kappa/2$, where $\kappa$ is the epicyclic frequency and $\Omega$ is the angular velocity of the disc rotation. Thus, it is expected that the OLR is the outer boundary of the spiral structure, and the ILR is associated with the beginning of the spirals or the ends of the bar (if present). The position of the outer resonance, according to this paper, is determined as the optical radius of the galaxy with an accuracy of $10$\%. To estimate $R_c$, the positions at which bright parts of the spiral change to faint ones or where the number of arms changes are also taken into account in these works. The complete list of expected morphological features that took place at resonances is provided in table~4 of~\citet{Elmegreen(1992)}.

\subsection{Puerari-Dottori method (\textit{P-D})}
The method was first applied by~\citet{Puerari97}. Its main idea is based on the existence of an age gradient of stars across the spiral arms, similar to the \textit{offset} method. The corotation radius is measured as the distance where the angular \textit{offset} profile changes its sign. This work is based on a Fourier analysis of the azimuthal profile of galaxies using both blue optical and infrared images. It indicates young and old stellar populations, correspondingly.  

\subsection{Spiral's width (\textit{width})}
The main idea of this method is based on investigating the azimuthal distribution of matter in the galactic disc. As mentioned earlier, the spiral pattern inside the corotation circle is supposed to rotate faster than the disc material, and vice versa in the outer region. Additionally, a significant concentration of matter is observed inside the corotation radius, on the outer edge of the arms, due to the presence of a shock wave. The same effect is assumed to occur on the inner side of the spiral in the external region of the disc. Therefore, the inner and outer regions should have asymmetry relative to the centre of mass in the azimuthal distribution of spirals, but the skew of their profiles has different signs. Near the corotation radius position, the distribution is supposed to be more symmetrical. Hence, the corotation resonance is estimated as the galactocentric distance where the skew profile changes sign. An illustrative explanation of the essence of this method is presented in fig.~1 in~\citet{Marchuk2024}.  

\end{document}